\documentclass[preprint]{aastex}

\received{2002 May 6}
\accepted{}
\slugcomment{To appear in The Astrophysical Journal}

\shorttitle{TIMESCALE FOR MASS SEGREGATION}
\shortauthors{MOURI \& TANIGUCHI}

\begin{document}

\title{MASS SEGREGATION IN STAR CLUSTERS: ANALYTIC ESTIMATION OF THE TIMESCALE}

\author{HIDEAKI MOURI}
\affil{Meteorological Research Institute, Nagamine 1-1, Tsukuba 305-0052, Japan; hmouri@mri-jma.go.jp}

\and

\author{YOSHIAKI TANIGUCHI}
\affil{Astronomical Institute, Graduate School of Science, Tohoku University, Aoba, Sendai 980-8578, Japan; tani@astr.tohoku.ac.jp}

\begin{abstract}
Mass segregation in a star cluster is studied in an analytical manner. We consider a two-component cluster, which consists of two types of stars with different masses. Plummer's model is used for the initial condition. We trace the overall behaviors of the probability distribution functions of the two components and obtain the timescale of mass segregation as a simple function of the cluster parameters. The result is used to discuss the origin of a black hole with mass of $\gtrsim 10^3\,M_{\sun}$ found in the starburst galaxy M82.
\end{abstract}

\keywords{stellar dynamics --- 
          galaxies: star clusters ---
          galaxies: starburst}

\notetoeditor{You may find the expression ``$10^1$'' etc. in our order-of-magnitude discussion. Please do not replace it with, e.g., ``10''.}

\notetoeditor{In equations, we have adopted the style that physical quantities are distinguishable from the numerical factor for the entire equation. Please do not change this style.}

\section{INTRODUCTION}

Within a star cluster, massive objects segregate into the cluster core as they lose kinetic energies to the less massive objects during approach toward energy equipartition. The result is an increase of the number density of the massive objects in the core. This phenomenon of mass segregation has long been known (Spitzer 1969; Binney \& Tremaine 1987, p. 531) but has attracted new interests in recent years. The large number density induces successive mergers of those massive objects, which could lead to the formation of an exotic object such as a very massive black hole.

The important parameter of mass segregation is its timescale. In order to explain an observation in terms of mass segregation, the timescale has to be less than the age of the star cluster.

However, mass segregation has been studied only for specific cases using numerical methods, i.e., Fokker-Planck, Monte Carlo, and $N$-body simulations. The exceptional analytical works that are available are those of Spitzer (1969) and Tremaine, Ostriker, \& Spitzer (1975). Spitzer (1969) considered an isothermal cluster of uniform density and estimated the timescale of mass segregation as the timescale of energy equipartition between stars of different masses. This model is too idealized for a star cluster, which is not isothermal or of uniform density. Tremaine et al. (1975) studied a decaying circular motion of an object in the singular isothermal sphere. This is originally a model for a motion of a globular cluster around the center of a galaxy, and it is not a model for random motions of stars in a cluster. Therefore, an analytical study of a more appropriate situation is desirable.

We analytically study a two-component cluster, which consists of two masses $m_0$ and $m$ ($m_0 < m$). The major component is the less massive stars. Their total mass $M_0$ dominates over the total mass $M$ of the more massive stars, $M_0 \gg M$. Here the subscript 0 is used to indicate that the quantity is relevant to the major component. The cluster is  spherically symmetric in all its properties and completely isotropic in the velocity space. 

Since the timescale of mass segregation is expected to be less than the relaxation timescale (Spitzer 1969), the star cluster is regarded as almost collisionless. The individual stars move almost freely in the mean potential. The gravitational encounters occur only in a stochastic manner. Our study is accordingly based on the overall behaviors of the probability distribution functions (PDFs) of the major and minor components. They are approximated by PDFs of completely collisionless systems. The minor-component PDF is from a parameterized family of steady-state PDFs for which the degree of mass segregation is allowed to change. The major-component PDF is kept the same (\S2). These PDFs are used to obtain the energy loss per unit time from the minor component due to encounters with the major component (\S3). The energy loss rate yields the rate of change of the minor-component PDF, which in turn yields the timescale of mass segregation (\S4, eq. [\ref{eq26}]). 

We compare our result with those for relevant timescales, including the pioneering work of Spitzer (1969). We also use our result to discuss the origin of a black hole with mass of $\gtrsim 10^3\,M_{\sun}$ found in a young star cluster of the starburst galaxy M82 (\S5).

\section{PROBABILITY DISTRIBUTION FUNCTIONS}

The major component, i.e., the less massive stars, is assumed to follow Plummer's model (polytrope with $n = 5$; Binney \& Tremaine 1987, p. 223). This steady-state solution for a collisionless spherical system has a simple analytic form but well reproduces observations of some globular clusters. The PDF in the phase space is determined by two parameters, e.g., the total mass $M_0$ and the core radius $r_{\rm c}$: 
\begin{equation}
\label{eq1}
f_0 = \frac{32\sqrt{2}}{7 \pi ^2}
      \frac{3 M_0}{4 \pi r_{\rm c}^3}
      \left( \frac{GM_0}{r_{\rm c}} \right) ^{-5}
      \left( \Psi_0 -\frac{v^2}{2}  \right) ^{7/2},
\end{equation}
with $f_0 = 0$ at $v \ge (2 \Psi_0)^{1/2}$. Here $G$ is the gravitational constant and $v$ is the velocity. The relative potential $\Psi_0$ is a function of the radius $r$ and defined as the negative of the gravitational potential $\Phi_0 \le 0$:
\begin{equation}
\label{eq2}
\Psi_0 (r) = -\Phi_0 (r)
           = \frac{GM_0}{r_{\rm c}}
             \frac{1}{\left( 1+r^2/r_{\rm c}^2 \right) ^{1/2}}.
\end{equation}
The corresponding mass density $\rho_0$ is
\begin{equation}
\label{eq3}
\rho_0 (r) = 4 \pi \int^{\infty}_0 v^2 f_0 dv
           = \frac{3M_0}{4 \pi r_{\rm c}^3}
             \frac{1}{\left( 1+r^2/r_{\rm c}^2 \right) ^{5/2}}.
\end{equation}
Here the integral ${\textstyle 4 \pi \int^{\infty}_0 r^2 \rho_0 dr}$ gives the total mass $M_0$. The squared velocity dispersion $\sigma _0 ^2$ is
\begin{equation}
\label{eq4}
\sigma _0 ^2 (r) = \frac{4 \pi \int^{\infty}_0 v^4 f_0 dv}
                        {4 \pi \int^{\infty}_0 v^2 f_0 dv}
                 = \frac{GM_0}{2r_{\rm c}}
                   \frac{1}{\left( 1+r^2/r_{\rm c}^2 \right) ^{1/2}}.
\end{equation}

\placefigure{Fig1}
\placefigure{Fig2}
\placetable{1}

The minor component, i.e., the more massive stars, is assumed to have no contribution to the gravitational field of the star cluster. Then the gravitational field is determined solely by the relative potential $\Psi_0$ of the major component. According to the Jeans theorem for a steady-state collisionless spherical system (Binney \& Tremaine 1987, p. 221), any function of $\Psi_0-v^2/2$ serves as the minor-component PDF. The PDF assumed in our study is from a function family that has a parameter $\beta \ge 1$:
\begin{equation}
\label{eq5}
f_{\beta} = \frac{32\sqrt{2}}{7 \pi^2} F_1 F_2
            \frac{3 M}{4 \pi r_{\rm c}^3}
            \left( \frac{GM_0}{r_{\rm c}} \right) ^{-(7\beta+3)/2}
            \left( \Psi_0 -\frac{v^2}{2}  \right) ^{7\beta/2},
\end{equation}
with $f_{\beta} = 0$ at $v \ge (2 \Psi_0)^{1/2}$. The factors $F_1$ and $F_2$ are functions of $\beta$ defined as
\begin{equation}
\label{eq6}
\int^{\infty}_0 \frac{x^2 dx}
                     {\left( 1+x^2 \right)^{(7\beta+3)/4}} 
=
\frac{1}{3}
\frac{1}{F_1(\beta)},
\end{equation}
and
\begin{equation}
\label{eq7}
\int^1_0 x^2 \left( 1-x^2 \right)^{7\beta/2} dx 
=
\frac{7\pi}{512}
\frac{1}{F_2(\beta)},
\end{equation}
where $F_1 = F_2 = 1$ at $\beta = 1$. Their numerical values are shown in Table 1 and Figure 1. The corresponding mass density $\rho_{\beta}$ is 
\begin{equation}
\label{eq8}
\rho_{\beta} (r) = F_1 \frac{3M}{4 \pi r_{\rm c}^3}
                       \frac{1}{\left( 1+r^2/r_{\rm c}^2 \right) ^{(7\beta+3)/4}},
\end{equation}
with the total mass $M$. The squared velocity dispersion $\sigma^2_{\beta}$ is
\begin{equation}
\label{eq9}
\sigma ^2_{\beta} (r) = \frac{12}{7\beta+5}
                        \frac{GM_0}{2r_{\rm c}}
                        \frac{1}{\left( 1+r^2/r_{\rm c}^2 \right) ^{1/2}}
                        \propto \sigma_0^2 (r).
\end{equation}
Figure 2 shows the radial profiles of the mass density and squared velocity dispersion for $\beta = 1$, 10, and 100. The total energy $E_{\beta}$ is obtained by integrating $\rho_{\beta} (\sigma^2_{\beta}+\Phi_0) = \rho_{\beta} (\sigma^2_{\beta}-\Psi_0)$ over the entire volume:
\begin{equation}
\label{eq10}
E_{\beta} = - \frac{3 \pi}{32}
              \frac{2(7\beta-1)}{7\beta+5} F_1 F_3 
              \frac{GM_0M}{r_{\rm c}}.
\end{equation}
The factor $F_3$ is a function of $\beta$ defined as
\begin{equation}
\label{eq11}
\int^{\infty}_{0} \frac{x^2 dx}
                       {\left( 1+x^2 \right)^{(7\beta+5)/4}} 
=
\frac{\pi}{16}F_3(\beta),
\end{equation}
where $F_3 = 1$ at $\beta = 1$. We define the initial state as $f_{\beta = 1}$ so that the radial profile of mass density and the velocity dispersion are the same as those of the major component. With an increase of the $\beta$ value, the mass density of the minor component becomes more centrally concentrated than that of the major component. Simultaneously, the velocity dispersion of the minor component becomes smaller (Fig. 2). These behaviors represent mass segregation. The radial profile of velocity dispersion of the minor component remains the same as that of the major component. This property is essential to establishing the final state of mass segregation where no net energy exchange exists between the two components (\S3). Thus, with a single parameter $\beta \ge 1$, the function family $f_{\beta}$ provides the simplest model to trace the evolution of the minor component from the initial to final states of mass segregation. At an intermediate state, the minor-component PDF is set to be just one of the functions, $f = f_{\beta}$, or a linear combination of them, $f = \sum (M_{\beta}/M) f_{\beta}$ with $M = \sum M_{\beta}$ (\S4).

\section{ENERGY CHANGE RATE}

Consider a single object in the minor component, which loses kinetic energy due to gravitational encounters  with objects in the major component. If the velocity of the single object is $v$, the rate of change of kinetic energy per unit mass is obtained as
\begin{equation}
\label{eq12}
\left\langle 
\frac{\partial}{\partial t} \left( \frac{v^2}{2} \right) 
\right\rangle
= 
16 \pi ^2 G^2 \ln \Lambda 
\left( 
m_0 \int^{\infty}_v v_0             f_0 dv_0 -
m   \int^v_0        \frac{v_0^2}{v} f_0 dv_0 
\right),
\end{equation}
with $f_0 = f_0(\Psi_0 - v_0^2/2)$ (Binney \& Tremaine 1987, eq. [8-66]). Here $\langle \cdot \rangle$ denotes an ensemble average. The cutoff factor $\Lambda$ is the ratio of the maximum and minimum impact parameters, $b_{{\rm max}}$ and $b_{{\rm min}}$:
\begin{equation}
\label{eq13}
\Lambda = \frac{b_{{\rm max}}}{b_{{\rm min}}}
        = \frac{b_{{\rm max}} v_{{\rm rel}}^2}{G(m+m_0)} \simeq \frac{m_0}{m+m_0}N_0.
\end{equation}
Here $v_{{\rm rel}}$ is a typical relative velocity between objects in the cluster (Binney \& Tremaine 1987, p. 511), and $N_0 = M_0/m_0$ is the total number of objects in the major component. We have assumed $v_{{\rm rel}}^2 \simeq G M_0 / b_{{\rm max}} = G m_0 N_0 / b_{{\rm max}}$.

By multiplying equation (\ref{eq12}) by the minor-component PDF $f$ and integrating over the $v$ space, we obtain the rate of change of kinetic energy per unit volume of the minor component. For the PDF $f = f_{\beta}(\Psi_0 - v^2/2)$, the result is
\begin{eqnarray}
\label{eq14}
\left\langle
\frac{\partial}{\partial t} 
\left( 
\frac{\rho \sigma^2}{2} 
\right) 
\right\rangle
& = & 
- \frac{2^{19} \sqrt{2}}{49 \pi} 
  \frac{16!!}{19!!}  
  F_1 F_2 F_4
%%%%%%%%%%%%%%%%%%%%%%%%%
% \nonumber \\ & & \times
%%%%%%%%%%%%%%%%%%%%%%%%%
  \frac{1}{ \left( 1+r^2/r_{\rm c}^2 \right)^{(7\beta+12)/4} }
\nonumber \\ & & \times
  G^2 \ln \Lambda
  \left( 
  \frac{m}{7\beta+2} - \frac{m_0}{9} 
  \right)
%%%%%%%%%%%%%%%%%%%%%%%%%
% \nonumber \\ & & \times
%%%%%%%%%%%%%%%%%%%%%%%%%
  \frac{3M_0}{4 \pi r_{\rm c}^3}
  \frac{3M}  {4 \pi r_{\rm c}^3}
  \left( 
  \frac{GM_0}{r_{\rm c}} 
  \right) ^{-1/2}.
\end{eqnarray}
The factor $F_4$ is a function of $\beta$ defined as 
\begin{equation}
\label{eq15}
\int^{1}_{0} x^2 \left( 1-x^2 \right) ^{(7\beta+9)/2} dx 
= 
\frac{16!!}{19!!} F_4(\beta),
\end{equation}
where $F_4 = 1$ at $\beta = 1$. By integrating equation (\ref{eq14}) over the entire volume, we obtain the rate of change of total energy $E$ of the minor component:
\begin{eqnarray}
\label{eq16}
\frac{dE}{dt} 
&=&
- \frac{2^{21} \sqrt{2 \pi}}{3 \times 5 \times 7^2}
  \frac{16!!}{19!!}
  \frac{\Gamma (13/4)}{\Gamma (15/4)}
  F_1 F_2 F_4 F_5 
  r_{\rm c}^3
\nonumber \\ & & \times
  G^2 \ln \Lambda
  \left( 
  \frac{m}{7\beta+2} - \frac{m_0}{9} 
  \right)
%%%%%%%%%%%%%%%%%%%%%%%%%
% \nonumber \\ & & \times
%%%%%%%%%%%%%%%%%%%%%%%%%
  \frac{3M_0}{4 \pi r_{\rm c}^3}
  \frac{3M}  {4 \pi r_{\rm c}^3}
  \left( 
  \frac{GM_0}{r_{\rm c}} 
  \right) ^{-1/2}.
\end{eqnarray}
The factor $F_5$ is a function of $\beta$ defined as
\begin{equation}
\label{eq17}
\int^{\infty}_{0} \frac{x^2 dx}
                       { \left( 1+x^2 \right) ^{(7\beta+12)/4}}
=
\frac{\sqrt{\pi}}{15}
\frac{\Gamma (13/4)}{\Gamma (15/4)}
F_5 (\beta),
\end{equation}
where $F_5 = 1$ at $\beta = 1$.

Recall that gravitational encounters in a star cluster are stochastic. The energy change rates (\ref{eq12}) and (\ref{eq14}) have been obtained as ensemble averages. They are not locally defined quantities. The energy change rate for the entire minor component (\ref{eq16}) has been obtained by assuming $\langle dE/dt \rangle = dE/dt$ because the cluster contains many stars.

The first and second integrals in equation (\ref{eq12}) represent heating and cooling by the major component, respectively. They yield the terms $m_0 /9$ and $m/(7\beta +2)$, respectively, in equations (\ref{eq14}) and (\ref{eq16}). When the two terms balance, the total energy of the minor component settles to an equilibrium, $dE/dt = 0$. Hence the $\beta$ value corresponding to the final state of mass segregation is
\begin{equation}
\label{eq18}
\beta _{{\rm fin}} = \frac{9m-2m_0}{7m_0}.
\end{equation}
This relation does not imply equipartition of kinetic energies, $m \sigma^2 = m_0 \sigma_0^2$ (see eqs. [\ref{eq4}] and [\ref{eq9}]), because the major component is kept to follow Plummer's nonthermal model. The condition for energy equipartition in an equilibrium state is that each of the two components has a thermal velocity distribution (Chapman \& Cowling 1970, p. 80). A qualitatively consistent result for very large values of $m/m_0$ was obtained by Chatterjee, Hernquist, \& Loeb (2002).\footnote{
Chatterjee et al. obtained the squared velocity dispersion $\sigma^2$ at $r \rightarrow 0$ for a Brownian motion of a single massive object that is in equilibrium with the background stars. Their study is mathematically equivalent to ours because we have ignored encounters among objects in the minor component. The background stars were assumed to follow Plummer's model as in our present study.}

\section{TIMESCALE OF MASS SEGREGATION}

By using the energy change rate $dE/dt$ in equation (\ref{eq16}), we estimate the timescale of mass segregation $\tau _{{\rm ms}}$. This timescale is defined in terms of the secular increase of the mass of the minor component, i.e., the more massive stars, in the cluster core $M_{\rm c}$:
\begin{equation}
\label{eq19}
\tau _{{\rm ms}} = \left. \frac{M_{\rm c}}{dM_{\rm c}/dt} \right| _{t=0},
\end{equation}
with 
\begin{equation}
\label{eq20}
M_{\rm c} = 4 \pi \int^{r_{\rm c}}_0 r^2 \rho dr 
          = \frac{F_1 F_6}{2 \sqrt{2}} M,
\end{equation} 
for $f = f_{\beta}$. The factor $F_6$ is a function of $\beta$ defined as
\begin{equation}
\label{eq21}
\int^1_0 \frac{x^2 dx}{\left( 1+x^2 \right) ^{(7\beta+3)/4}}
=
\frac{1}{6 \sqrt{2}} F_6 (\beta),
\end{equation}
where $F_6 = 1$ at $\beta = 1$.

Since it is not analytically determined how the minor-component PDF changes in the course of mass segregation, we study two extreme cases. The one is that the minor-component PDF is characterized by a single $\beta$ value, which increases from the initial value of 1 toward the final value $\beta_{{\rm fin}}$, i.e., $f = f_{\beta}$ with $\beta = 1$ at $t = 0$ ($1 \le \beta \le \beta_{{\rm fin}}$). Equations (\ref{eq6}), (\ref{eq20}), and (\ref{eq21}) yield $dM_{\rm c}/d\beta$, while equations (\ref{eq6}), (\ref{eq10}), and (\ref{eq11}) yield $dE/d\beta$. Then $dM_{\rm c}/dt$ is obtained as $(dM_{\rm c}/d\beta)(d\beta/dt) = (dM_{\rm c}/d\beta)(dE/dt)/(dE/d\beta)$. Using their initial values, we obtain
\begin{eqnarray}
\label{eq22}
\tau _{{\rm ms}}
& = &
\frac{3^3 \times 5 \times 7^2 \pi \sqrt{\pi}}{2^{23} \sqrt{2}}
\frac{19!!}{16!!}
\frac{\Gamma (15/4)}{\Gamma (13/4)} 
%%%%%%%%%%%%%%%%%%%%%%%%%
% \nonumber \\ & & \times
%%%%%%%%%%%%%%%%%%%%%%%%%
\frac{7-8\ln2}
     {8\ln (1+\sqrt{2})-3\sqrt{2} \ln2 -2\sqrt{2}}
\nonumber \\ & & \times
\frac{1}{G^2 \ln \Lambda}
\frac{1}{m-m_0}
\left( 
\frac{3M_0}{4 \pi r_{\rm c}^3} 
\right) ^{-1}
\left( 
\frac{GM_0}{2r_{\rm c}} 
\right) ^{3/2}.
\end{eqnarray}
The numerical factor is about 0.3878. 

The other extreme case is that the state $\beta = 1$ evolves directly into the state $\beta = \beta_{{\rm fin}}$. These two states serve as the bases for the minor-component PDF, i.e., $f = (M_{\beta = 1}/M) f_{\beta =1} + (M_{\beta = \beta_{{\rm fin}}}/M) f_{\beta = \beta_{{\rm fin}}} = (1- \gamma ) f_{\beta = 1} + \gamma f_{\beta = \beta_{{\rm fin}}}$ with $\gamma = 0$ at $t = 0$ ($0 \le \gamma \le 1$). The core mass $M_{\rm c}$ and the total energy $E$ of the minor component are
\begin{equation}
\label{eq23}
M_{\rm c} = (1-\gamma) \frac{1}{2 \sqrt{2}}M + \gamma M,
\end{equation}
and
\begin{equation}
\label{eq24}
E = - (1-\gamma) \frac{3 \pi}{32} \frac{GM_0M}{r_{\rm c}} 
    - \gamma \frac{GM_0M}{r_{\rm c}} .
\end{equation}
Here we have made approximations for the state $\beta = \beta_{{\rm fin}}$ taking account of the fact that the minor component is well concentrated in the core and has a negligible velocity dispersion. Then $dM_{\rm c}/dt$ is obtained as $(dM_{\rm c}/d\gamma)(d\gamma/dt) = (dM_{\rm c}/d\gamma)(dE/dt)/(dE/d\gamma)$. Using their initial values, we obtain
\begin{eqnarray}
\label{eq25}
\tau _{{\rm ms}}
& = &
\frac{3^2 \times 5 \times 7^2 \sqrt{\pi}}{2^{23}}
\frac{19!!}{16!!}
\frac{\Gamma (15/4)}{\Gamma (13/4)}
\frac{32-3\pi}
     {2\sqrt{2}-1}
\nonumber \\ & & \times
\frac{1}{G^2 \ln \Lambda}
\frac{1}{m-m_0}
\left( 
\frac{3M_0}{4 \pi r_{\rm c}^3} 
\right) ^{-1}
\left( 
\frac{GM_0}{2r_{\rm c}} 
\right) ^{3/2}.
\end{eqnarray}
The numerical factor is about 0.6331. The present case has a somewhat larger timescale than the previous case because the effective energy difference is somewhat larger, i.e., $dE/d\gamma < dE/d\beta < 0$ while $dM_{\rm c}/d\gamma \simeq dM_{\rm c}/d\beta$ in the initial state.

Mass segregation in a real star cluster is expected to be intermediate between the above two cases. According to Fokker-Planck and Monte Carlo simulations of early stages of mass segregation (Inagaki \& Wiyanto 1984; Fregeau et al. 2002), the ratio of velocity dispersions $\sigma / \sigma _0$ averaged over the whole cluster does not change appreciably, while that averaged over the core alone approaches steadily to the final value. The reason is that gravitational encounters occur preferentially among stars on orbits that pass through the dense core. Although this fact has been ignored in our present calculation, the behavior of $\sigma / \sigma _0$ would be reproduced if the minor-component PDF were a superposition of various states between those of $\beta = 1$ and $\beta = \beta_{{\rm fin}}$ with the average $\beta$ value increasing with the time.

Equations (\ref{eq22}) and (\ref{eq25}) have the same dependence on physical quantities. The values of the numerical factors are also similar to each other and well approximated to be $0.5 \pm 0.1$ or $2^{-1}$. Thus the timescale of mass segregation is
\begin{equation}
\label{eq26}
\tau _{{\rm ms}} = 
\frac{1}{2}
\frac{1}{G^2 \ln \Lambda}
\frac{1}{m-m_0}
\left( 
\frac{3M_0}{4 \pi r_{\rm c}^3} 
\right) ^{-1}
\left( 
\frac{GM_0}{2r_{\rm c}} 
\right) ^{3/2},
\end{equation}
where $3M_0/4 \pi r_{\rm c}^3$ and $(G M_0 /2r_{\rm c})^{1/2}$ correspond to the central mass density $\rho_0(0)$ and central velocity dispersion $\sigma_0(0)$ of the major component, respectively (eqs. [\ref{eq3}] and [\ref{eq4}]).

\section{DISCUSSION}
\subsection{Comparison with Relevant Timescales}

The timescale of mass segregation is greater than the core crossing timescale. Since the core radius is $r_{\rm c}$ and the central velocity dispersion is $(GM_0/2r_{\rm c})^{1/2}$, the core crossing timescale $\tau _{{\rm cc}}$ is
\begin{equation}
\label{eq27}
\tau _{{\rm cc}} = \left( \frac{2r_{\rm c}^3}{GM_0} \right) ^{1/2}.
\end{equation}
Then equation (\ref{eq26}) for the timescale of mass segregation $\tau _{{\rm ms}}$ is rewritten as
\begin{equation}
\label{eq28}
\tau _{{\rm ms}}
=
\frac{\pi}{6} 
\frac{m_0 N_0 /(m-m_0)}{\ln \left[ m_0 N_0 /(m+m_0) \right]} 
\tau _{{\rm cc}}
\gg
\tau _{{\rm cc}},
\end{equation}
where we have assumed $m_0/m \simeq 10^{-1}$--$10^{-2}$ and $N_0 \simeq 10^5$--$10^6$ as a practical situation. On the other hand, the timescale of mass segregation is less than the relaxation timescale $\tau_{{\rm rlx}}$ of the major component, which is about $N_0 / 8\ln N_0$ times the crossing timescale $\tau_{{\rm cc}}$ (Binney \& Tremaine 1987, p. 190):
\begin{equation}
\label{eq29}
\tau_{{\rm rlx}} \simeq \frac{1}{8} \frac{N_0}{\ln N_0} \tau_{{\rm cc}} > \tau_{{\rm ms}}.
\end{equation}
These results justifies our assumption that the star cluster is almost collisionless and the major-component PDF remains the same during mass segregation.

The timescale of mass segregation is close to the timescale of energy loss from the minor component, $\tau_{{\rm el}} = E/(dE/dt)$ at $t=0$. Using equations (\ref{eq10}) and (\ref{eq16}), we obtain
\begin{eqnarray}
\label{eq30}
\tau_{{\rm el}} &=& 
\frac{3^3 \times 5 \times 7^2 \pi \sqrt{\pi}}{2^{23}}
\frac{19!!}{16!!}
\frac{\Gamma (15/4)}{\Gamma (13/4)}
%%%%%%%%%%%%%%%%%%%%%%%%%%%
% \nonumber \\ & & \times 
%%%%%%%%%%%%%%%%%%%%%%%%%%%
\frac{1}{G^2 \ln \Lambda}
\frac{1}{m-m_0}
\left( 
\frac{3M_0}{4 \pi r_{\rm c}^3} 
\right) ^{-1}
\left( 
\frac{GM_0}{2r_{\rm c}} 
\right) ^{3/2}.
\end{eqnarray}
The dependence on physical quantities is the same as that of equation (\ref{eq26}). This is because the energy loss from the minor component and the increase of its central concentration are of the same phenomenon. The numerical factor of equation (\ref{eq30}) is about 0.4832, which happens to be in agreement with the numerical factor 2$^{-1}$ of equation (\ref{eq26}). Since this energy loss timescale is obtained without any assumption on the evolution of the minor-component PDF, it is a robust measure.

\subsection{Comparison with Spitzer (1969)}

Spitzer (1969) considered an isothermal cluster of uniform density and estimated the timescale of mass segregation as the timescale of approach toward energy equipartition between the major and minor components,\footnote{
Defined as the time it takes for the difference between $m \sigma^2$ and $m_0 \sigma_0^2$ to decrease by a factor of $e^{-1}$.} 
which was about $m_0/m$ times the relaxation timescale $\tau_{{\rm rlx}}$ of the major component. Using equation (\ref{eq29}), we obtain the corresponding timescale in our model cluster:
\begin{equation}
\label{eq31}
\tau_{{\rm ee\,(Spitzer)}} 
\simeq
\frac{1}{8} 
\frac{m_0N_0/m}{\ln N_0} \tau_{{\rm cc}}.
\end{equation}
If $m \gg m_0$, this energy equipartition timescale is by definition equivalent to our energy loss timescale $\tau_{{\rm el}}$ in equation (\ref{eq30}). Since the numerical factor is $2^{-1}$, the equation is rewritten as
\begin{equation}
\label{eq32}
\tau _{{\rm el}}
=
\frac{\pi}{6} \frac{m_0 N_0 /m}{\ln \left( m_0 N_0 /m \right)} \tau _{{\rm cc}},
\end{equation}
where we have replaced $m \pm m_0$ with $m$. Except for an unimportant difference in the argument of the logarithm, the timescales (\ref{eq31}) and (\ref{eq32}) have the same dependence on physical quantities. On the other hand, the numerical factors are different. This difference is not serious because there is ambiguity in the definition of the relaxation timescale. Our relaxation timescale (\ref{eq29}) is based on the core crossing timescale $\tau_{{\rm cc}}$ and might not be sufficiently large to yield the energy equipartition timescale for the whole cluster. Overall, although Spitzer (1969) considered a very idealized cluster, his timescale is expected to be valid for energy equipartition in a real star cluster at least as an order-of-magnitude estimation. This is also the case for mass segregation, which occurs during approach toward energy equipartition.

Spitzer (1969) also predicted that, if the total mass of the minor component is much less than that of the major component, $M \ll M_0$, the two components eventually achieve energy equipartition, $m \sigma^2 = m_0 \sigma_0^2$. His prediction has been confirmed at least for the cluster core by Fokker-Planck and Monte Carlo simulations (Inagaki \& Wiyanto 1984; Watters, Joshi, \& Rasio 2000: Fregeau et al. 2002). Since it is required that both the major and minor components evolve to have thermal velocity distributions (\S3), the energy equipartition is achieved at $t \gtrsim \tau_{{\rm rlx}}$.\footnote{
This energy equipartition in the cluster core is not exact. The difference between $m \sigma^2$ and $m_0 \sigma_0^2$ is very small but still existent ($\ga 1$\%; Inagaki \& Wiyanto 1984; Watters et al. 2000). The reason is that the velocity distributions do not become exactly thermal even in the cluster core.} 
This is beyond the limit of our model ($t < \tau_{{\rm rlx}})$, where the cluster is assumed to be almost collisionless. The major component is kept to follow Plummer's nonthermal model and hence does not achieve energy equipartition with the minor component, regardless of the total mass ratio $M/M_0$. This fact is not serious to our estimation of the timescale of mass segregation, which is less than the relaxation timescale $\tau_{{\rm rlx}}$ and has been defined at the initial time of the evolution $t = 0$ in equation (\ref{eq19}).

\subsection{Application to Observations}
If the central mass density $\rho_0(0)$ and central velocity dispersion $\sigma_0(0)$ have values typical of globular clusters (Binney \& Tremaine 1987, p. 26, see also p. 423 for the cutoff factor $\Lambda$), equation (\ref{eq26}) is written as
\begin{equation}
\label{eq33}
\tau _{{\rm ms}}
=     
3 \times 10^7\,{\rm yr}
\left( 
\frac{\ln \Lambda}{10} 
\right) ^{-1}
\left( 
\frac{m}{10\,M_{\sun}} 
\right) ^{-1}
\left( 
\frac{\rho_0(0)}{10^4\,M_{\sun}\,{\rm pc}^{-3}} 
\right) ^{-1}
\left( 
\frac{\sigma_0(0)}{10\,{\rm km}\,{\rm s}^{-1}}  
\right) ^3.
\end{equation}
Here we have assumed $m \gg m_0$ and replaced $m-m_0$ with $m$. If we instead use the typical values of the core radius $r_{\rm c}$ and central velocity dispersion $\sigma_0(0)$, both of which are observable quantities, equation (\ref{eq26}) is rewritten as
\begin{equation}
\label{eq34}
\tau _{{\rm ms}}
=
2 \times 10^7\,{\rm yr}
\left( 
\frac{\ln \Lambda}{10} 
\right) ^{-1}
\left( 
\frac{m}{10\,M_{\sun}} 
\right) ^{-1}
\left( 
\frac{r_{\rm c}}{1\,{\rm pc}} 
\right) ^2
\left( 
\frac{\sigma_0(0)}{10\,{\rm km}\,{\rm s}^{-1}}  
\right).
\end{equation}
Thus mass segregation in a typical star cluster has the timescale $\tau_{{\rm ms}}$ of order $10^7$ yr.

The above timescale of mass segregation is used to discuss the origin of a very massive black hole found in the starburst galaxy M82 (Kaaret et al. 2001; Matsumoto et al. 2001; see also Matsushita et al. 2000). At the 2 \micron\ secondary peak, i.e., an active site of star formation, there is a source of compact X-ray emission. The observed strong variability implies that the source is an accreting black hole. The observed luminosity of $10^{41}$\,ergs\,s$^{-1}$ implies that the mass is greater than $10^3$\,$M_{\sun}$ if the emission is isotropic and its luminosity is below the Eddington limit. This black hole should have been formed via successive mergers of massive stars and black holes that had segregated into the core of a star cluster (Taniguchi et al. 2000). Since the 2 \micron\ secondary peak has a starburst age of about $1 \times 10^7$\,yr (Satyapal et al. 1997), the mass segregation should have occurred well within this short duration. The timescales given in equations (\ref{eq33}) and (\ref{eq34}) are too long. We suspect that the star cluster is exceptionally dense and compact. For example, the young star cluster R136 in the Large Magellanic Cloud has $\rho_0(0) \simeq 10^6\,M_{\sun}$\,pc$^{-3}$, $\sigma_0(0) \simeq 10^1$\,km\,s$^{-1}$, and $r_{\rm c} \simeq 10^{-2}$--$10^{-1}$\,pc (Portegies Zwart et al. 1999), which yields $\tau _{{\rm ms}} \simeq 10^5$\,yr. In fact, only one black hole heavier than $10^3\,M_{\sun}$ has been found among more than 100 young star clusters of M82. The objects that had segregated into the cluster core should have merged in a runaway manner (Quinlan \& Shapiro 1989; Portegies Zwart et al. 1999; Mouri \& Taniguchi 2002) because the merger probability is higher for more massive objects.

\acknowledgments
This work has been supported in part by the Japanese Ministry of Education, Science, and Culture under grants 10044052 and 10304013. We are grateful to N. Kanaeda and the referee for helpful comments.

\begin{deluxetable}{rllllll}

\tablenum{1}
\tablecolumns{7}
\tablewidth{0pc}
\tablecaption{NUMERICAL VALUES OF THE FACTORS $F_1$--$F_6$}

\tablehead{
\colhead{$\beta$} &
\colhead{$F_1$} &
\colhead{$F_2$} &
\colhead{$F_3$} &
\colhead{$F_4$} &
\colhead{$F_5$} &
\colhead{$F_6$}}

\startdata
  1 & \phn\phn\phn 1.000 & \phn\phn 1.000 & 1.000     & 1.000    & 1.000    & 1.000    \\
  2 & \phn\phn\phn 3.875 & \phn\phn 2.294 & 0.3469    & 0.6177   & 0.5424   & 0.5480   \\
  3 & \phn\phn\phn 7.761 & \phn\phn 3.902 & 0.1881    & 0.4292   & 0.3519   & 0.3325   \\
  5 & \phn\phn     17.81 & \phn\phn 7.868 & 0.08728   & 0.2508   & 0.1914   & 0.1573   \\
 10 & \phn\phn     52.71 & \phn     21.15 & 0.03083   & 0.1081   & 0.07705  & 0.05366  \\
 20 & \phn         152.4 & \phn     58.30 & 0.01090   & 0.04263  & 0.02921  & 0.01856  \\
 30 & \phn         282.1 &          106.2 & 0.005933  & 0.02412  & 0.01628  & 0.01003  \\
 50 & \phn         610.4 &          226.8 & 0.002757  & 0.01157  & 0.007717 & 0.004633 \\
100 &              1734. &          638.1 & 0.0009748 & 0.004189 & 0.002769 & 0.001631 \\
\enddata

\end{deluxetable}

%\clearpage
\section*{FIGURE CAPTIONS}

\figcaption[Fig1]{{\it Solid lines:} Numerical values of $F_1$, $F_2$, $F_3$, $F_4$, $F_5$, and $F_6$. {\it Dotted lines:} Numerical values of $F_1 F_2$, $F_1 F_3$, $F_1 F_2 F_4$, $F_1 F_2 F_4 F_5$, and $F_1 F_6$. The abscissa is the parameter $\beta$.}

\figcaption[Fig2]{{\it Solid lines:} Radial profiles of the mass density $\rho_{\beta}$ in equation (\ref{eq8}) for $\beta = 1$, 10, and 100. {\it Dotted lines:} Radial profiles of the squared velocity dispersion $\sigma_{\beta}^2$ in equation (\ref{eq9}). The abscissa is the radius $r$ normalized by the core radius $r_{\rm c}$. For convenience, we have assumed $3M/4 \pi r_{\rm c}^3 = GM_0/2 r_{\rm c} = 1$.}

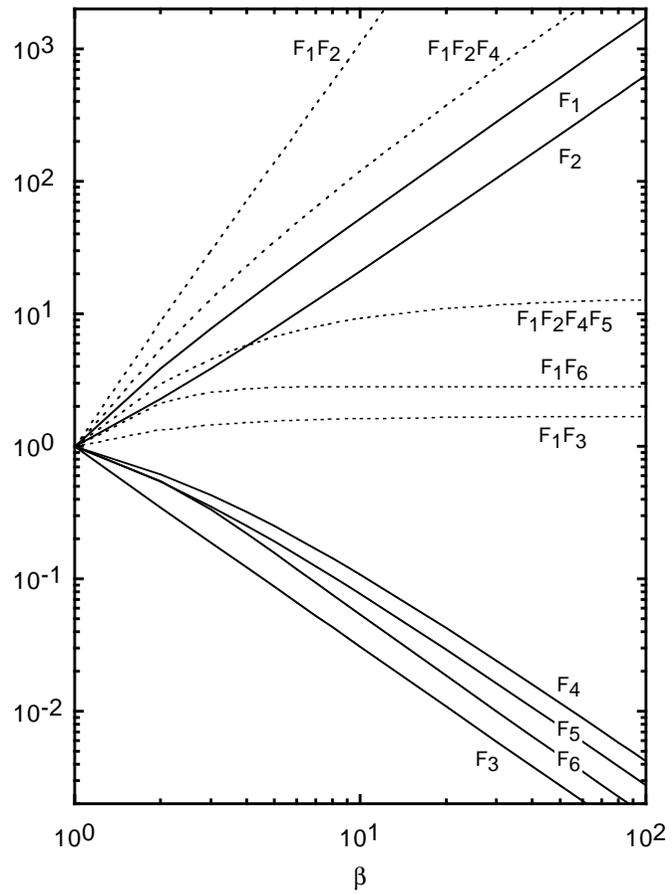
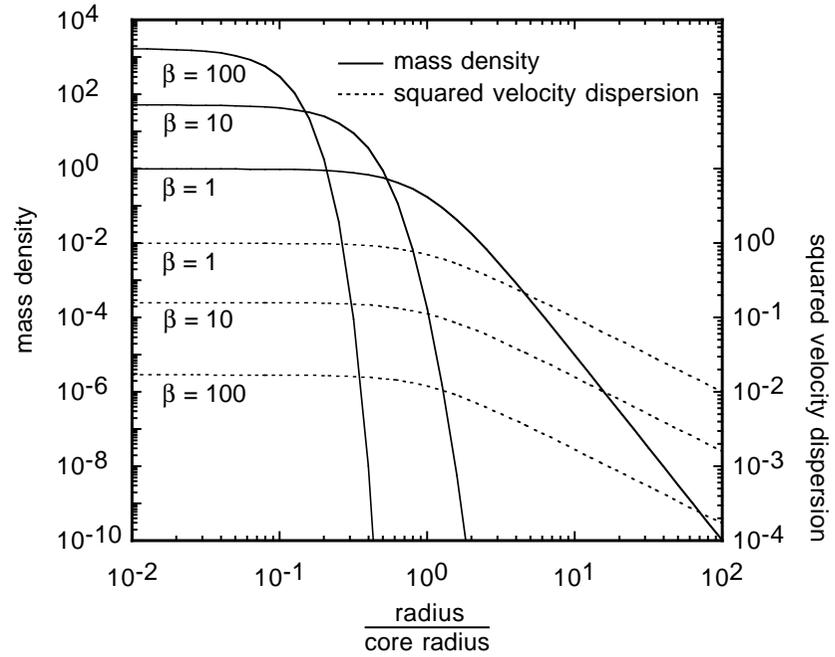
\end{document}